\documentclass[a4paper,showpacs,preprintnumbers,showkeys,twocolumn]{revtex4-1}

\usepackage{amsmath,amssymb,textcomp,float,dsfont}
\usepackage[utf8]{inputenc}
\usepackage{amsmath}
\usepackage{amsfonts}
\usepackage{amssymb,multirow}
\usepackage{times,fullpage}
\usepackage{comment,graphicx}
\usepackage{array}

\begin{document}
\newcommand{\nwc}{\newcommand}
\nwc{\vs}{\vspace}
\nwc{\hs}{\hspace}
\nwc{\la}{\langle}
\nwc{\ra}{\rangle}
\nwc{\nn}{\nonumber}
\nwc{\Ra}{\Rightarrow}
\nwc{\wt}{\widetilde}
\nwc{\lw}{\linewidth}
\nwc{\ft}{\frametitle}
\nwc{\ben}{\begin{enumerate}}
\nwc{\een}{\end{enumerate}}
\nwc{\bit}{\begin{itemize}}
\nwc{\eit}{\end{itemize}}
\nwc{\dg}{\dagger}
\nwc{\mA}{\mathcal A}
\nwc{\mD}{\mathcal D}
\nwc{\mB}{\mathcal B}

\nwc{\Tr}[1]{\underset{#1}{\mbox{Tr}}~}
\nwc{\pd}[2]{\frac{\partial #1}{\partial #2}}
\nwc{\ppd}[2]{\frac{\partial^2 #1}{\partial #2^2}}
\nwc{\fd}[2]{\frac{\delta #1}{\delta #2}}
\nwc{\pr}[2]{K(i_{#1},\alpha_{#1}|i_{#2},\alpha_{#2})}
\nwc{\av}[1]{\left< #1\right>}

\nwc{\zprl}[3]{Phys. Rev. Lett. ~{\bf #1},~#2~(#3)}
\nwc{\zpre}[3]{Phys. Rev. E ~{\bf #1},~#2~(#3)}
\nwc{\zpra}[3]{Phys. Rev. A ~{\bf #1},~#2~(#3)}
\nwc{\zjsm}[3]{J. Stat. Mech. ~{\bf #1},~#2~(#3)}
\nwc{\zepjb}[3]{Eur. Phys. J. B ~{\bf #1},~#2~(#3)}
\nwc{\zrmp}[3]{Rev. Mod. Phys. ~{\bf #1},~#2~(#3)}
\nwc{\zepl}[3]{Europhys. Lett. ~{\bf #1},~#2~(#3)}
\nwc{\zjsp}[3]{J. Stat. Phys. ~{\bf #1},~#2~(#3)}
\nwc{\zptps}[3]{Prog. Theor. Phys. Suppl. ~{\bf #1},~#2~(#3)}
\nwc{\zpt}[3]{Physics Today ~{\bf #1},~#2~(#3)}
\nwc{\zap}[3]{Adv. Phys. ~{\bf #1},~#2~(#3)}
\nwc{\zjpcm}[3]{J. Phys. Condens. Matter ~{\bf #1},~#2~(#3)}
\nwc{\zjpa}[3]{J. Phys. A: Math theor  ~{\bf #1},~#2~(#3)}

\title{Universal fluctuations in orbital diamagnetism: A surprise in theoretical physics}

\author{P. S. Pal$^{1,2}$\email{}, Arnab Saha$^3$\email{} and A. M. Jayannavar$^{1,2}$ \email{}}

\email{priyo@iopb.res.in, sahaarn@gmail.com, jayan@iopb.res.in}
\affiliation{\vspace{0.5cm}$^1$ Institute of Physics,  Sachivalaya Marg, Bhubaneswar-751005, India\\
$^2$ Homi Bhabha National Institute, Training School Complex, Anushakti Nagar, Mumbai-400085, India\\
$^3$ Savitribai Phule Pune University, Ganeshkhind, Pune-411007, India}

\begin{abstract}
 Over the last century Bohr van Leuween theorem attracted the notice of physicists. The theorem states about the absence of magnetization in classical systems in thermal equilibrium. In this paper, 
 we discuss about fluctuations of magnetic moment in classical systems. In recent years this topic has been investigated intensively and it is not free from controversy. We a have considered 
 a system consisting of a single particle moving in a plane. A magnetic field is applied perpendicular  to the plane. The system is in contact with a thermal bath. We have considered three cases:
 (a) particle moving in a homogeneous medium, (b) particle moving in a medium with space dependent friction  and (c) particle moving in a medium with space dependent temperature. For all the three 
 cases average magnetic moment and fluctuations in magnetic moment has been calculated. Average magnetic  moment saturates to a finite value in case of free particle but goes to zero when the particle
 is confined by a 2-D harmonic potential. Fluctuations in magnetic moment shows universal features in the presence of arbitrary friction inhomogeneity. For this case the system reaches equilibrium
 asymptotically. In case of space dependent temperature profile, the stationary distribution is non-Gibbsian and fluctuations deviate from universal value for the bounded system only. 
\end{abstract}
\pacs{05.10.Gg, 05.40.-a, 75.20.-g, 75.47.-m}
\keywords{diamagnetism, fluctuation, magnetic field}
\maketitle

\section{Introduction}
The behavior of free electrons in an external magnetic field ($B$) has long been of interest. Due to the circular motion of the electrons one would expect non-vanishing magnetic moment even without 
considering the spin of the electrons. A naive approach would lead us to find the radius of the orbit of the electron as $r=\frac{mv}{eB}$, where $v$ is the velocity of the electron and $B$ is the 
applied magnetic field. Hence the magnetic moment ($M$) of the orbit should be $evr$ or 
\begin{equation}
 M=\frac{mv^2}{B}.
\end{equation}
This implies that as $B\rightarrow 0$, the magnetic moment diverges. This mistake arises from the consideration of complete orbits. The above argument is due to Pierls\cite{pierls79}, who considers this 
problem as a surprise in theoretical physics. This error has been discussed by Neils Bohr and H. J. van Leuween
separately in their PhD dissertation almost a century ago. They had shown that in presence of constant magnetic field and in thermal equilibrium, the magnetization of an electron gas in the classical 
Drude-Lorentz (DL) model is identically zero.  This is known as {\it Bohr-van Leuween} (BvL) theorem\cite{leuween}. Over many  decades a lot of eminent physicists have been puzzled by this in the sense that 
physically it seems strange. Mathematically, one can deduce this result from the fact that the free energy, when calculated from the canonical partition function, is independent of the external magnetic field.
The average magnetic moment, being the derivative  of the free energy with respect to the magnetic field, is identically zero.

\textbf{A simple proof}: Hamiltonian of a classical particle of mass $m$ and charge
$e$ in an electromagnetic field, 

\begin{equation}
H(\vec r,\vec p)=\frac{1}{2m}\left[\vec p-\frac{e}{c}\vec A\right]^2+e\phi,
\end{equation}
where $\vec A$ is the vector potential and $\phi$ is the electrostatic potential.
N-particle classical Hamiltonian

\begin{eqnarray}
H_{cl}(\{\vec r_i\}_{i=1}^N,\{\vec p_i\}_{i=1}^N)&=&\frac{1}{2m}\sum_{i=1}^N\left[\vec
p_i-\frac{e}{c}\vec A(\vec r_i)\right]^2\nn\\
&&+V(\{\vec r_i\}_{i=1}^N).
\end{eqnarray}
Partition function of the system is given by 
\begin{equation}
Z_d=\int \prod_{i=1}^N d^N\vec{r_i} d^N\vec{p_i} \exp(-\beta H_{cl}).\nn
\end{equation}
For each $\vec p_i$ integral we can make the change of variable as $\left(\vec
p_i-\frac{e}{c}\vec A\right)\rightarrow \vec p_i$, so that magnetic field no longer
appears in the partition function. Therefore free energy $F_{cl}=-k_BT\ln Z_{cl}$ is
independent of magnetic field. As a result magnetization defined as 
\begin{equation}
M=-\lim_{B\rightarrow 0}\frac{\partial F_{cl}}{\partial B} \nn
\end{equation}
 vanishes identically.
 Physically, this null result derives from the exact cancellation of the
orbital diamagnetic moment associated with the complete cyclotron orbits of the charged particles by the paramagnetic moment subtended by the incomplete orbits skipping the boundary in the opposite
sense. This physical picture clearly shows that boundary of the system plays an important role in the perfect cancellation of diamagnetic contribution arising from the bulk and paramagnetic 
contribution arising from cuspidal orbits at the boundary. If one considers an {\it unbound} system, it can lead to a non-zero magnetic moment\cite{jayan81, jayan_thesis} and hence a violation of BvL theorem. 

Based on this intuitive picture,  in \cite{nkumar09}, the authors have considered a finite system consisting of a particle moving on the surface of a shere in the presence of external magnetic field. 
Following a real space-time approach based on the classical Langevin equation, these authors have computed the orbital magnetic moment that gives a non-zero value and has the diamagnetic sign. This 
work has been questioned in \cite{mahanti09} and \cite{pradhan10}. The authors in \cite{mahanti09} had shown the non-existence of classical diamagnetism for a system consisting of a particle moving in a ring 
and subjected to external magnetic field. In \cite{pradhan10}, the authors pointed out that the classical Langevin dynamics for a charged particle on a closed curved surface in a time-independent magnetic field 
leads to the canonical distribution in the long time limit. Thus the BvL theorem holds even for a finite system without any boundary  and the average magnetic moment is zero.

In contrary to these recent works, in a much earlier work \cite{jayan81}, the authors had reported the presence(absence) of classical diamagnetism for an unbound(bound) system  governed by classical Langevin
dynamics. They have derived an  exact expression of average magnetic moment for an unbounded system consisting of charged particle of mass $m$ moving in a medium of friction coefficient $\gamma$ :
\begin{equation}
 \la M(t)\ra_{t\rightarrow\infty}=-\frac{|e|}{mc}\frac{k_BT\omega_c}{(\omega_c^2+\omega_r^2)},
 \label{mean}
\end{equation}
where $\omega_c=\frac{|e|B}{mc}$ and $\omega_r=\frac{\gamma}{m}$. In case of a harmonically bound system, the average magnetic moment vanishes at large time limit - in consensus with the BvL theorem\cite{jayan81}.
In the same work\cite{jayan81}, for a charged Brownian particle undergoing birth-death process, they have shown that it can exhibit classical diamagnetism. It is important in  indirect gap 
semiconductors, where electron-hole  pair  production-recombination requires phonons and hence depends sensitively on temperature and, of course, on compensation, the above condition may be 
realisable. 

 At equilibrium, fluctuations in orbital magnetic moment for a bound system drops exponentially with the mean being zero\cite{roy08}:
\begin{equation}
 P(M)=\frac{1}{2\mu_B}\left(\frac{\hbar \omega_0}{k_BT}\right)\exp\left(-\frac{\hbar \omega_0}{k_BT}\frac{|M|}{\mu_B}\right),
 \label{distribution}
\end{equation}
where, $\mu_B=e\hbar/2mc$ is the Bohr magneton and $\omega_0=\sqrt{k_0/m}$. The absence of diamagnetism has been shown using Jarzynski equality\cite{jar97,jar97_1,jar11,saha08}. When a system is 
subjected to external driving, the system can exhibit paramagnetism or diamagnetism in a non-equilibrium steady state depending on the physical parameters\cite{sahoo07,roy08,nkumar09,kumar12,nkumar12}. In the present work we
are interested in in fluctuations of orbital diamagnetism. To our surprise, we obtain universal fluctuations independent of the nature of the system. For this we have considered a system consisting 
of a single particle moving in a plane. A magnetic field is applied perpendicular  to the plane. The system is in contact with a thermal bath. We have considered three cases:
(a) particle moving in a homogeneous medium, (b) particle moving in a medium with space dependent friction  and (c) particle moving in a medium with space dependent temperature. 
In Section \ref{model}, we describe our model. In Section \ref{numerical} and \ref{results} we give the numerical details and discuss the results obtained from simulation both for unbounded
and bounded  system. Finally we conclude in Section \ref{conclusion}.
\section{Our model}
\label{model}
We consider a system consisting of a charged Brownian particle of mass $m$ and charge $e$ constrained to move on a two dimensional (X-Y plane) medium under the influence of a  two dimensional harmonic 
potential and a constant magnetic field $\vec{B}=B\hat{z}$ perpendicular to that plane. The  whole system is in contact with a heat bath at temperature $T$. We have considered three different cases: 
(a) particle moving in a homogeneous medium, (b) particle moving in a medium with space dependent friction and (c) particle moving in a medium with space dependent temperature.
Langevin equation for systems with space dependent friction has been derived from first principles in \cite{mahato96,jayan95}. We have chosen a symmetric gauge producing a constant magnetic 
field $B$ along $z$-direction. The dynamics of the Brownian particle is modeled by the following underdamped Langevin equation 
\begin{eqnarray}
 m\ddot x&=&-\gamma(x)\dot x-kx+eB\dot y+\sqrt{2k_BT(x)\gamma(x)}\eta_x(t),\nn\\
 m\ddot y&=&-\gamma(y)\dot y-ky-eB\dot x+\sqrt{2k_BT(y)\gamma(y)}\eta_y(t).\nn
\end{eqnarray}
Here $\eta_x$ and $\eta_y$ are the components of the thermal noise from the bath in $x$ and $y$ directions. The mean value of the Gaussian noise is zero and they are delta correlated with 
$\langle \eta_i(t')\eta_j(t'')\rangle=\delta_{ij}\delta(t'-t'')$  for $i,j=x,y$. The strength of the noise $D$, friction coefficient $\gamma(x)$ and temperature $T$ of the bath are related to each
other by the usual fluctuation dissipation relation, i.e., $D=\gamma(x) k_B T/m$, where $k_B$ is the Boltzmann constant.  However, fluctuation dissipation theorem is not valid in the presence of
space dependent temperature, which will be discussed in the next paragraph.

The dynamical evolution of a Brownian particle in an inhomogeneous medium with spatially varying friction and temperature field is important to understand conceptually. It requires to address the 
basic problem of relative stability of states in nonequilibrium systems which has been a subject of debate for over several decades. The theoretical treatments adopted so far are mostly
phenomenological in nature. Landauer, in a series of papers\cite{landauer78,landauer79,landauer83,landauer88,landauer93}, argues that for systems with nonuniform temperature the relative stability of two states will be affected by the detailed kinetics
all along the pathways (on the potential surface) between the two states under comparison. It is the effect of thermal fluctuations that plays a crucial role and the resulting effective
potential surface may have completely different nature from that with uniform temperature. With the help of his ``blowtorch'' theorem Landauer shows that a change of temperature away from 
uniformity even at very unlikely positions of the system on the potential surface may cause probability currents to set in moving the system towards a new steady state situation changing thereby 
the relative stability of the otherwise locally stable states. 

The variation of friction coefficient in space  changes the dynamics of the particle in the a potential field but eventually the system, which is in contact with a bath at fixed temperature,
approaches towards its equilibrium Boltzmann distribution. 
The relative stability of the competing states is generally governed by the usual Boltzmann factor in the local neighborhood of the corresponding (representative) potential wells. A change in the
potential barrier between two potential well minima changes the relaxation rate but leaves the relative stability of the two well-states unchanged.

\section{Numerical Simulation}
\label{numerical}
In this section, we focus on numerical results obtained by evolving the system using discretised Langevin dynamics with time step dt= 0.001 in the underdamped regime. 
The medium in which the particle is moving is considered to be inhomogeneous. Inhomogeneity arises in two different ways: 1. either friction coefficient ($\gamma$) is space dependent or 2. temperature
is space dependent. We considered three different types of space dependency both for friction and temperature : (A) cosine, (B) symmetric tanh and (C) asymmetric tanh. We kept the 
temperature(friction) to be constant when friction (temperature) is varying. The mass of the particle $m$ and Boltzmann constant $k_B$ is set to unity. All the parameters are in  dimensionless 
form. The temperature $T$  is taken to be 0.5 in presence of space dependent friction and the friction coefficient $\gamma$ is fixed to unity when temperature is varying.

\section{Results and discussions}
\label{results}

\subsection{Unbound system}
\begin{figure}[H]
\begin{centering}
\includegraphics[height=5 cm,width=5.5 cm]{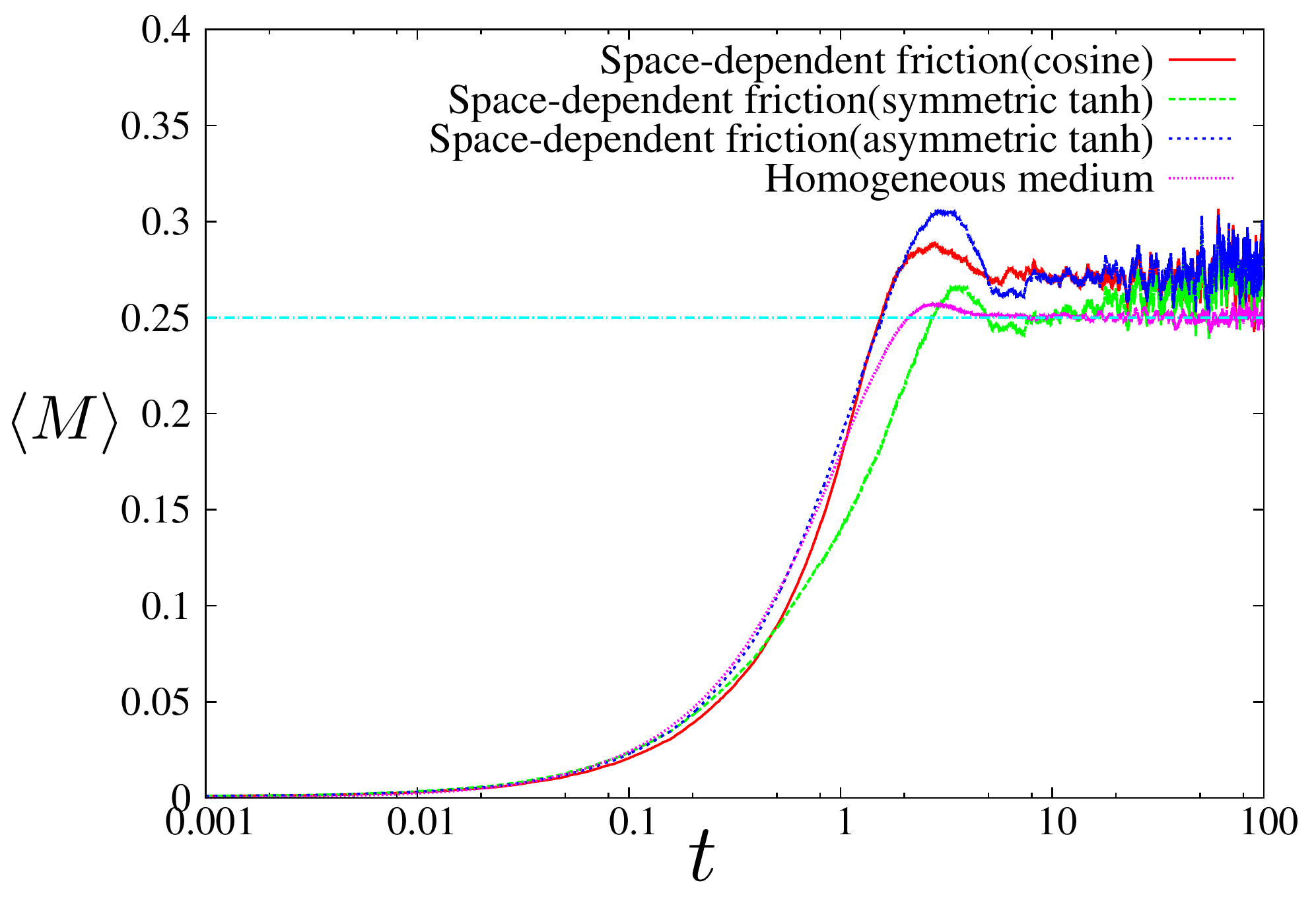}
\caption{(Color online) Average magnetic moment for unbounded system with space dependent friction.}
\label{avg_moment_vary_g_k0_log}
 \end{centering}
\end{figure}

\begin{figure}[H]
\begin{centering}
\includegraphics[height=5.5 cm,width=5.5cm]{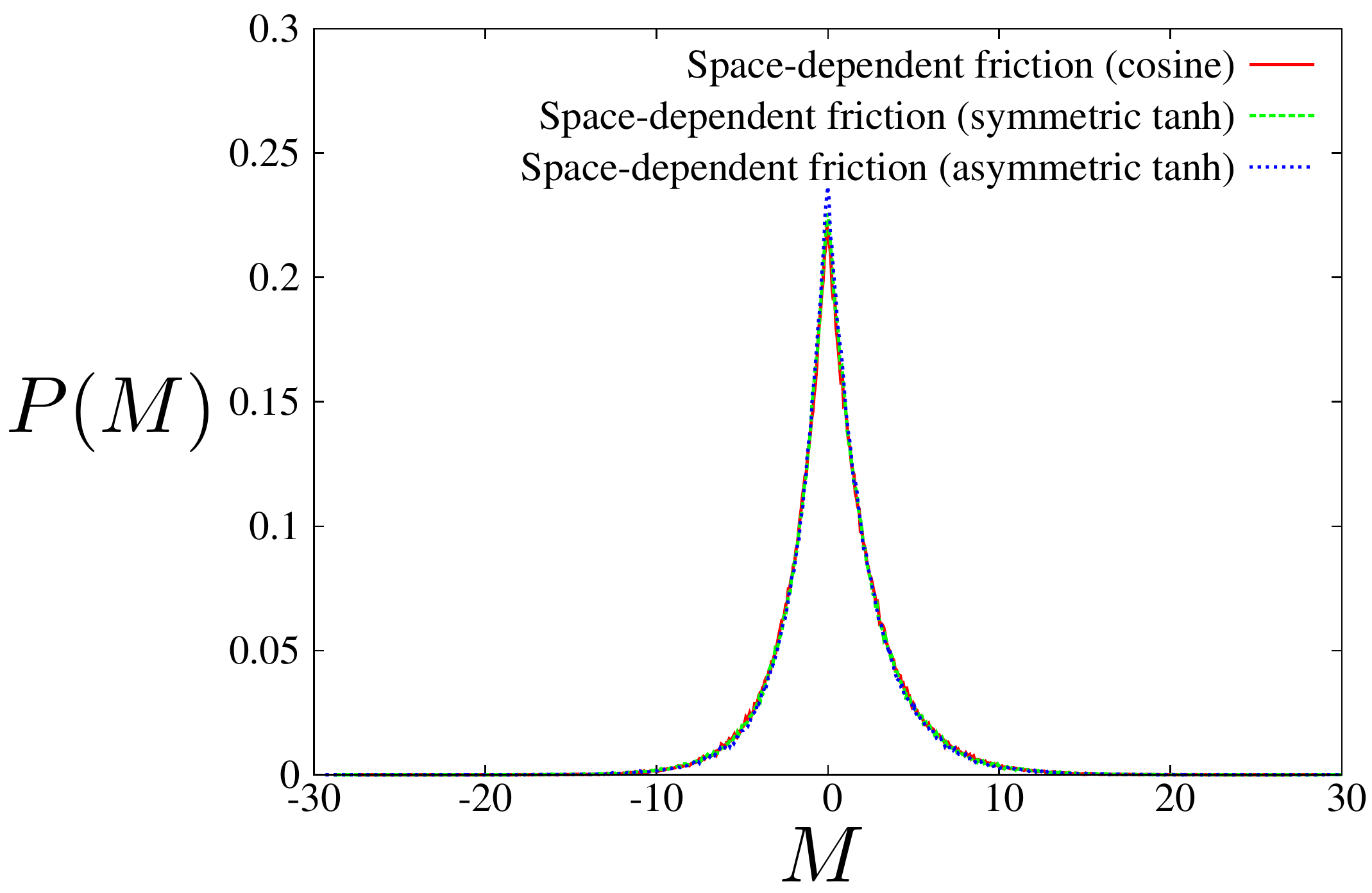}
\caption{(Color online) Probability distribution of orbital magnetic moment for unbounded system with space dependent friction.}
\label{dist_k0_vary_g}
 \end{centering}
\end{figure}

\begin{figure}[H]
\begin{centering}
\includegraphics[height=5 cm,width=5.5 cm]{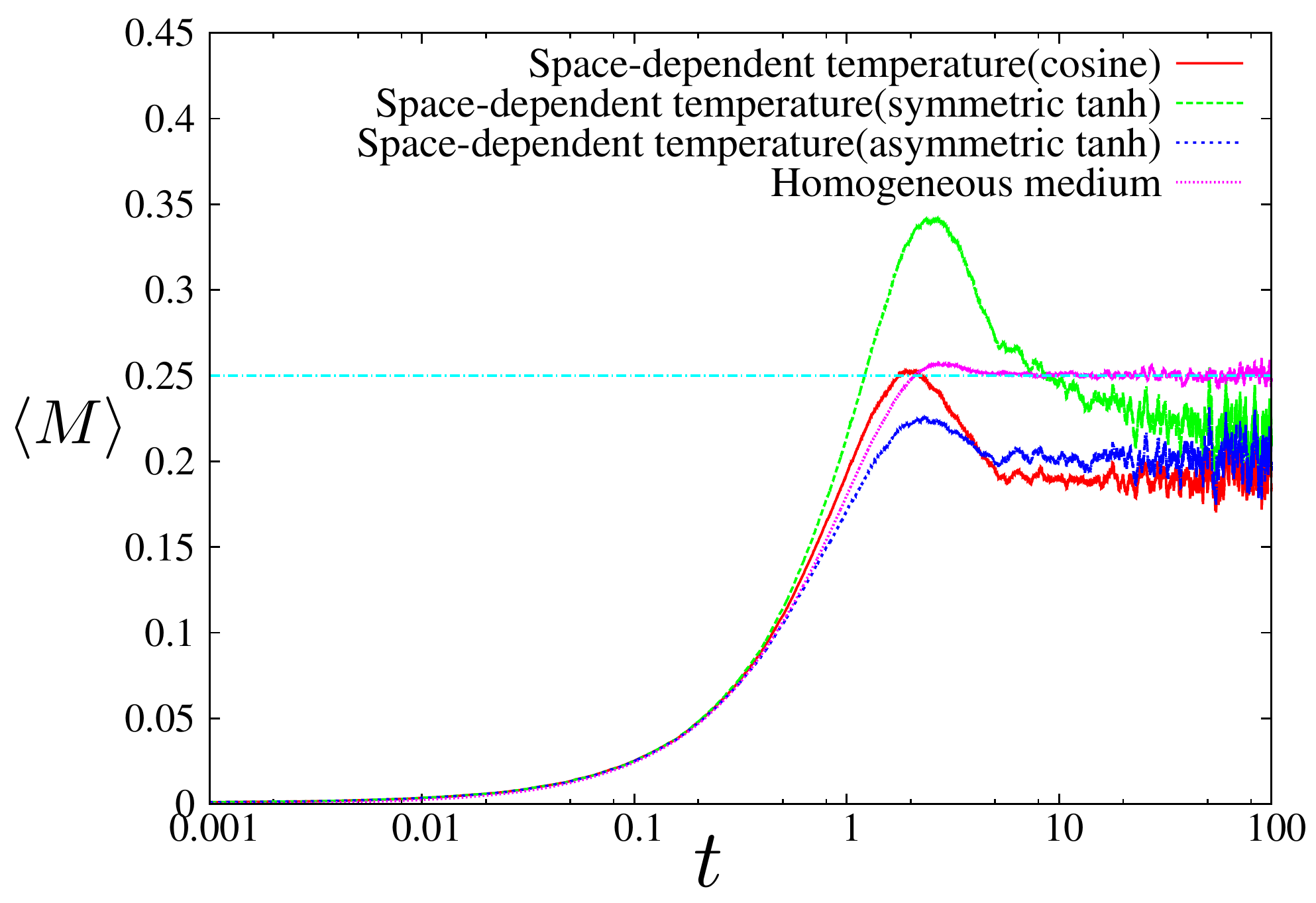}
\caption{(Color online) Average magnetic moment for unbounded system space dependent temperature.}
\label{avg_moment_vary_T_k0_log}
 \end{centering}
\end{figure}

\begin{figure}[H]
\begin{centering}
\includegraphics[height=5.5 cm,width=5.5cm]{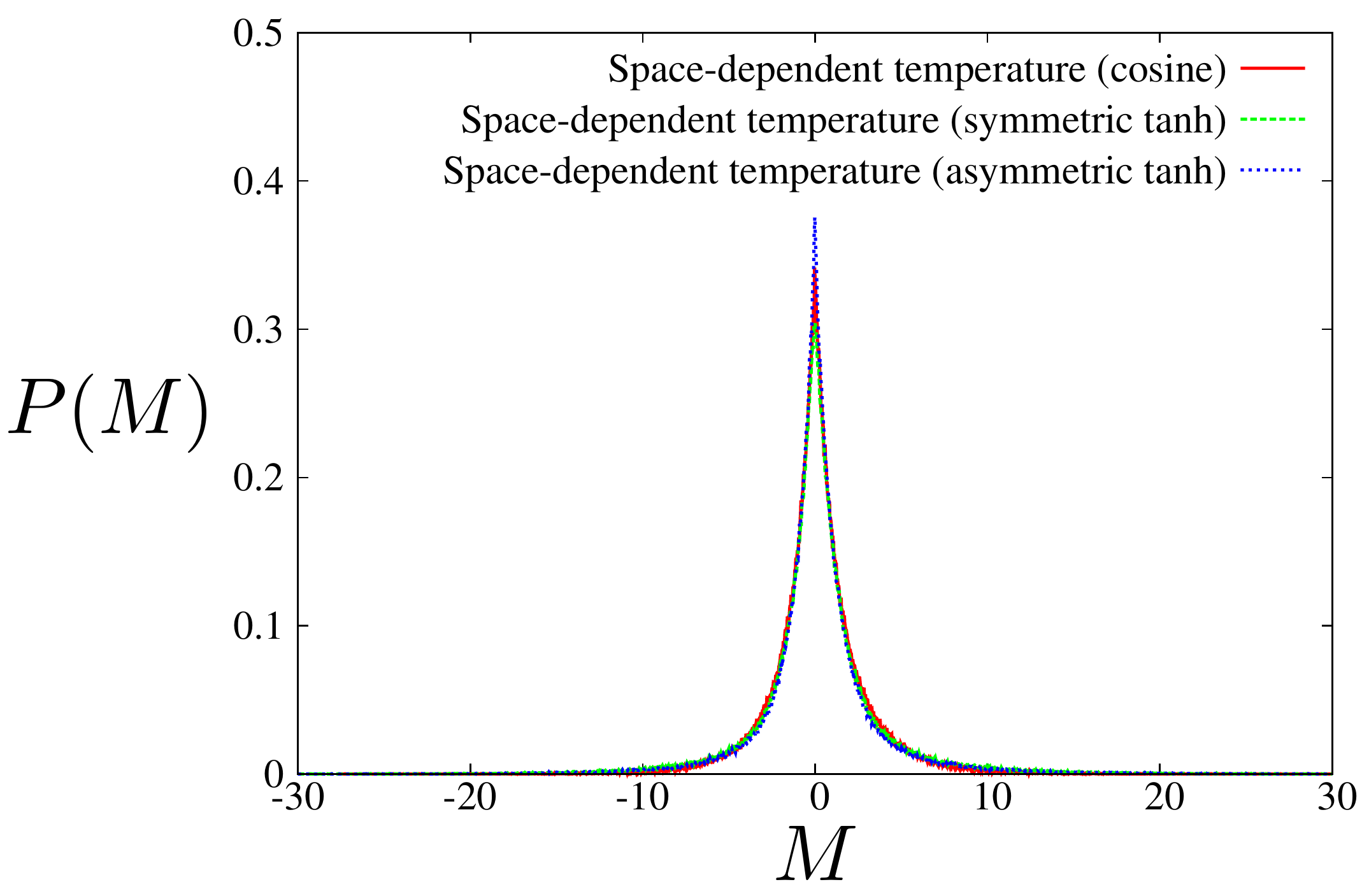}
\caption{(Color online) Probability distribution of orbital magnetic moment for unbounded system with space dependent temperature.}
\label{dist_k0_vary_T}
 \end{centering}
\end{figure}

In Fig.\ref{avg_moment_vary_g_k0_log}, we have plotted average magnetic moment $\la M\ra$ as a function of time for unbounded system for three forms of space dependent friction at constant temperature. The functional 
forms of space dependency of friction  are: 1. $\gamma(x)=\gamma_0(1-\lambda \cos(x/\gamma_1))$ with $\gamma_0=0.5, \lambda=0.9,\gamma_1=0.25$, 2. $\gamma(x)=\gamma_0+\gamma_1 \tanh[(x-\gamma_2)/\gamma_3]$ 
with $\gamma_0=0.5, \gamma_1=0.3, \gamma_2=0,\gamma_3=0.1$ and 3. $\gamma(x)=\gamma_0+\gamma_1 \tanh[(x-\gamma_2)/\gamma_3]$ with $\gamma_0=0.5, \gamma_1=0.3, \gamma_2=0.7,\gamma_3=0.1$. We 
notice that after some initial transients, which critically depends on the nature of functional form of friction coefficient, average magnetization asymptotically saturates to a constant value 0.25.
It is the same value given by Eq.\ref{mean}. 
In Fig.\ref{dist_k0_vary_g}, we have shown the fluctuations around the asymptotic value. We observe that these fluctuations are universal in nature. 
In case of space dependent temperature, we have kept the friction coefficient fixed to unity and considered same three functional form of space dependency as that of friction. Here we see from 
Fig.\ref{avg_moment_vary_T_k0_log} that average magnetic moment saturates at large time to a value which is different from that given by Eq.\ref{mean}. This is due to the fact that space dependent temperature drives the 
system out of equilibrium. Fig.\ref{dist_k0_vary_T} depicts the fluctuations of magnetic moment about the saturation value which clearly shows the universal behavior.
%
Table \ref{k0} summarizes the results for unbounded system.
\begin{table}[H]
\centering
\begin{tabular}{|c|c|c|c|}
\hline
 Variable & Space dependency & $\la M\ra $ & Theoretical value\\
 \hline
\multirow{3}{4em}{Friction ($\gamma$)} & Cosine & 0.27176 & \multirow{7}{4em}{0.25}\\
\cline{2-3}
&Symmetric tanh & 0.2671 &\\
\cline{2-3}
&Asymmetric tanh & 0.2758 &\\
\cline{1-3}
\multirow{3}{5em}{Temperature ($T$)} & Cosine & 0.4141 &\\
\cline{2-3}
&Symmetric tanh &  0.4502 &\\
\cline{2-3}
& Asymmetric tanh & 0.4331 &\\
\cline{1-3}
\multicolumn{2}{|c|}{Constant friction and temperature} & 0.2501 &\\
\hline
\end{tabular}
  \caption{For harmonically unbound system.}
 \label{k0}
\end{table}

\subsection{Bounded system}
\begin{figure}[H]
\begin{centering}
\includegraphics[height=5 cm,width=5.5 cm]{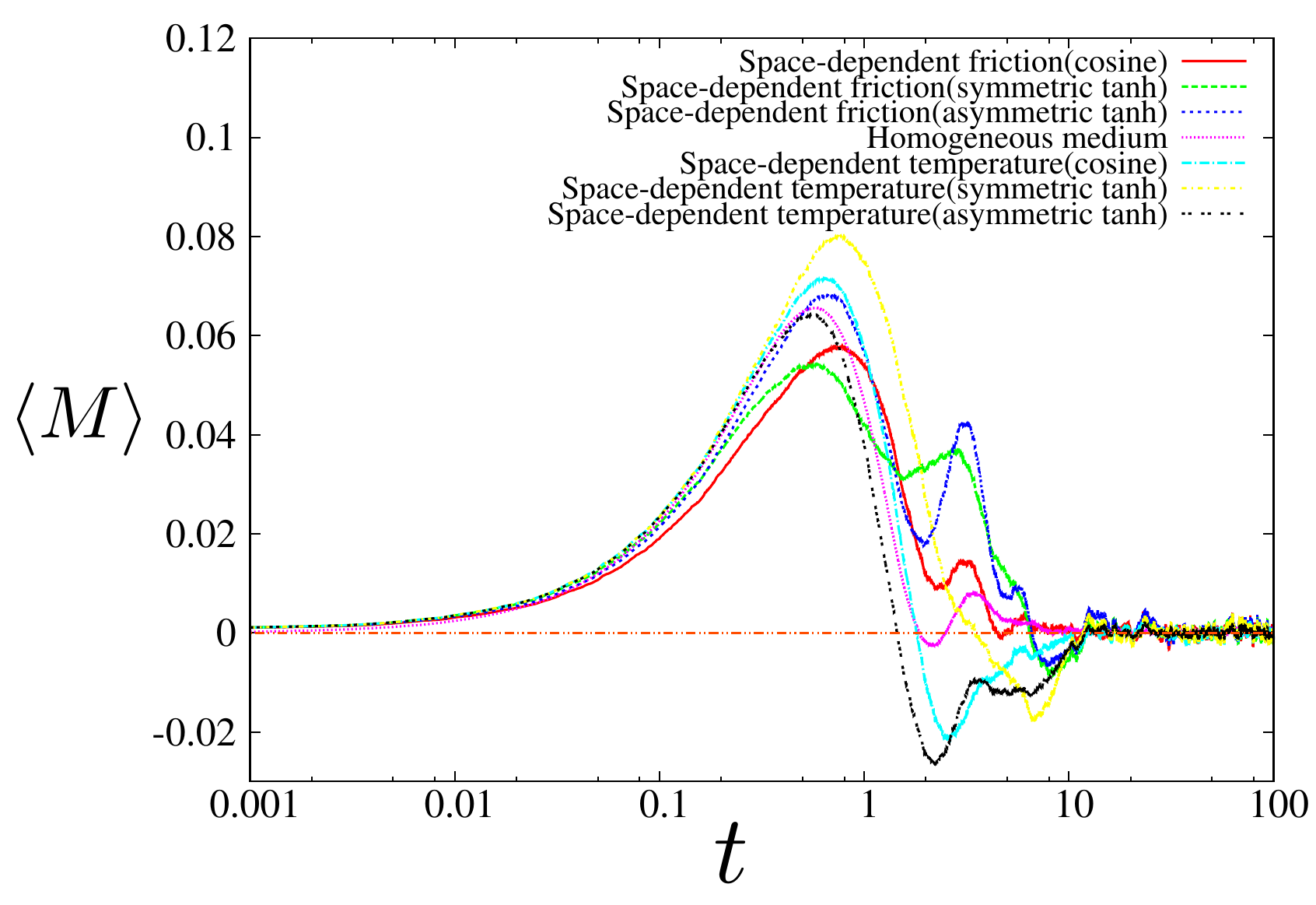}
\caption{(Color online) Average magnetic moment for bounded system.}
\label{mean_k1}
 \end{centering}
\end{figure}

\begin{figure}[H]
\begin{centering}
\includegraphics[height=5.5 cm,width=5.5cm]{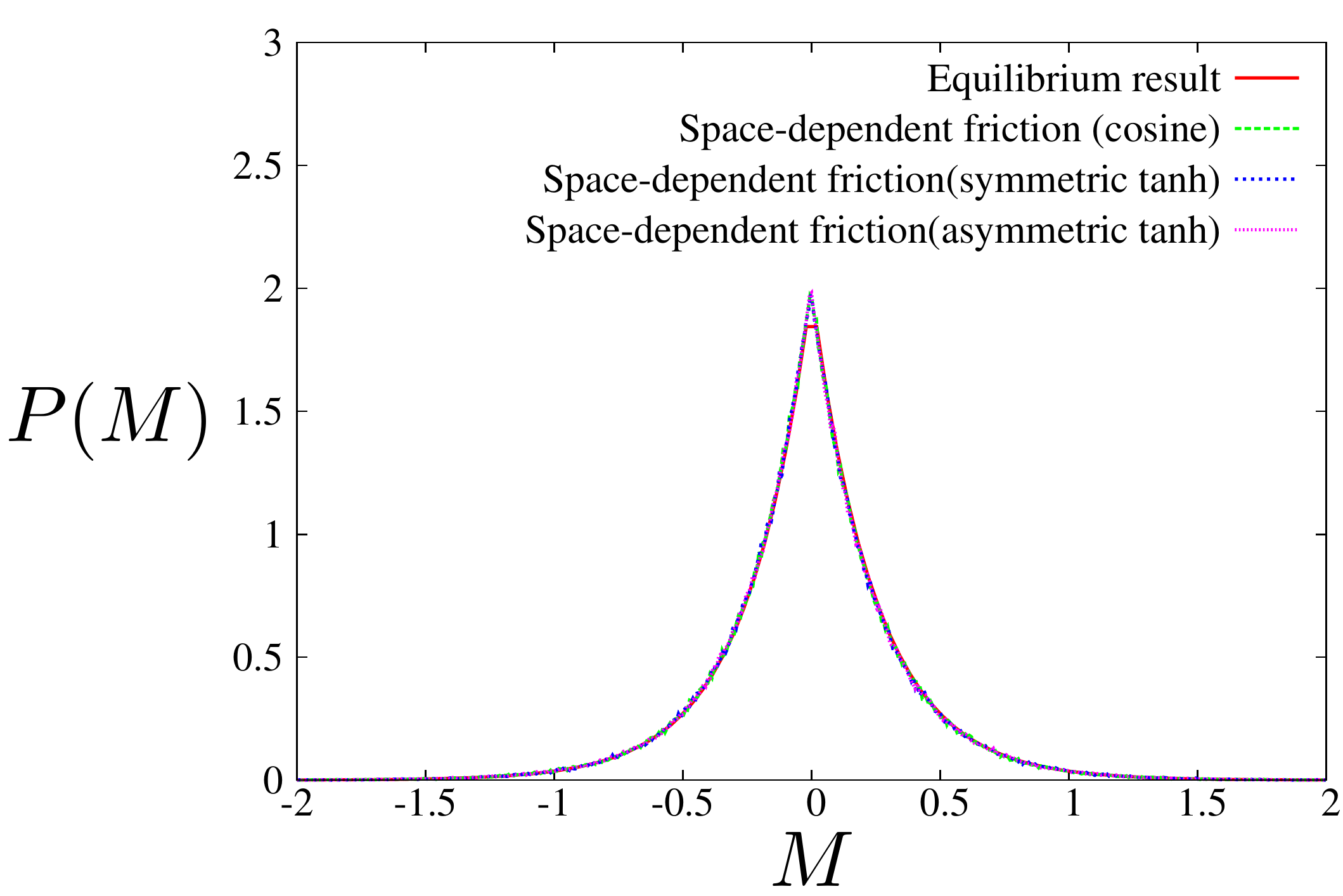}
\caption{(Color online) Probability distribution of orbital magnetic moment for bounded system with space dependent friction.}
\label{dist_k1_vary_g}
 \end{centering}
\end{figure}

In Fig.\ref{mean_k1}, we plotted the evolution of average orbital magnetic moment $\la M\ra$ for various space dependent friction and temperature. Again, we observe that after initial transients, depending
on inhomogeneity, average magnetization goes to zero asymptotically. Hence we recover BvL theorem irrespective of system inhomogeneity. 
In Fig.\ref{dist_k1_vary_g}, we have plotted probability distribution of magnetic moment $P(M)$ for different space dependent friction coefficient at constant temperature. To our surprise, we that $P(M)$ coincides
for  with the equilibrium result all the three cases.
\begin{figure}[H]
\begin{centering}
\includegraphics[height=5.5 cm,width=5.5cm]{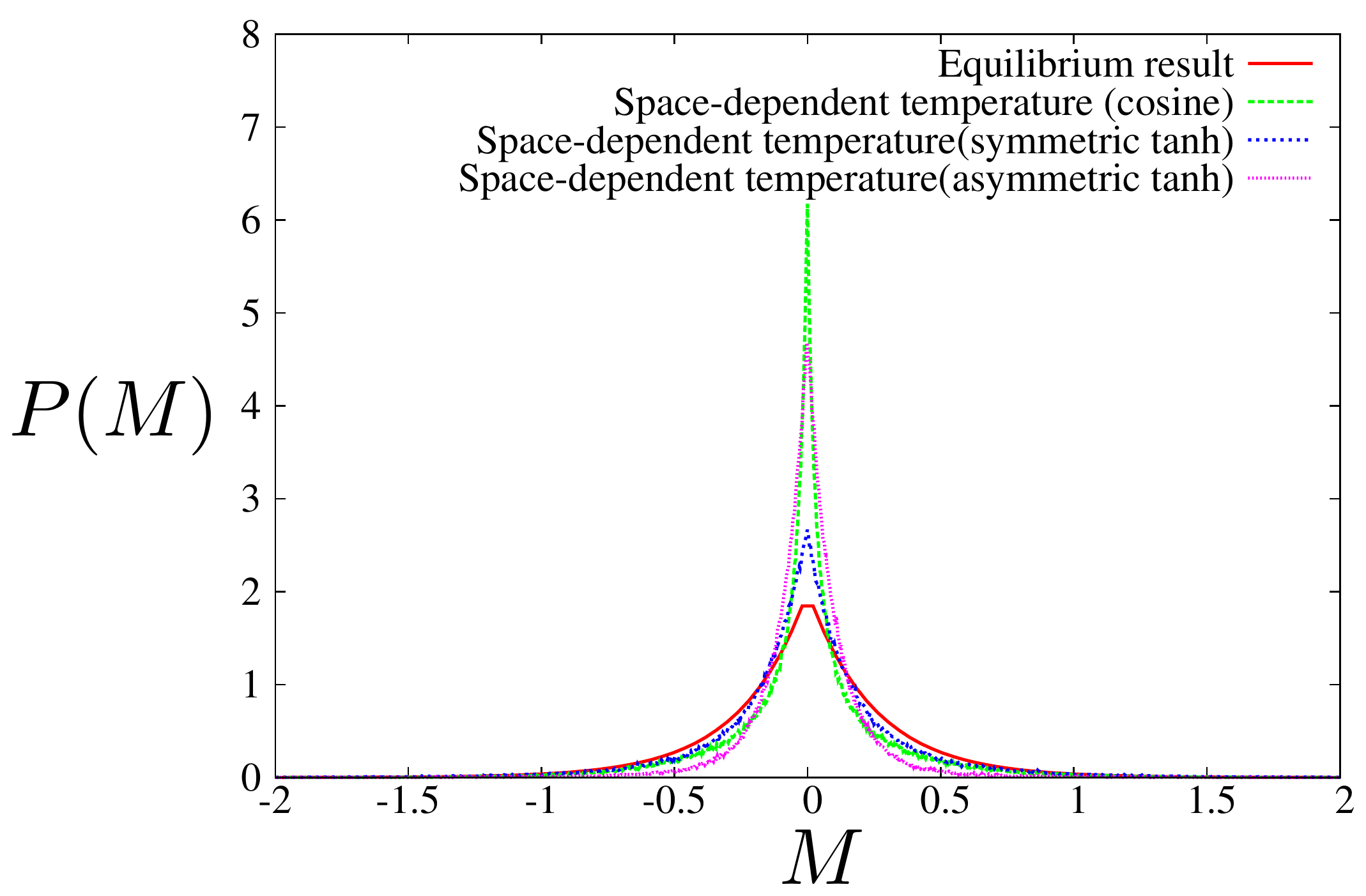}
\caption{(Color online) Probability distribution of orbital magnetic moment for bounded system with space dependent temperature.}
\label{dist_k1_vary_T}
 \end{centering}
\end{figure}
In Fig.\ref{dist_k1_vary_T}, we have plotted the fluctuations of magnetization around the asymptotic value keeping friction coefficient fixed. We observe that the system, being out of equilibrium, does not exhibit 
any universal behavior. A clear deviation in probability distribution $P(M)$ in case of space dependent temperature from that of the equilibrium system is also seen. 
Table \ref{k1} summarizes the results for bounded systems.

\begin{table}[H]
 \centering
\begin{tabular}{|c|c|c|c|c|}
\hline
 Variable & Space dependency & $\sigma_M $ & Theoretical value\\
 \hline
\multirow{3}{4em}{Friction ($\gamma$)} & Cosine & 0.3533 & \multirow{7}{4em}{0.354}\\
\cline{2-3}
&Symmetric tanh & 0.3535 &\\
\cline{2-3}
&Asymmetric tanh & 0.35385 &\\
\cline{1-3}
\multirow{3}{5em}{Temperature ($T$)} & Cosine & 0.6217 &\\
\cline{2-3}
&Symmetric tanh &  0.70892 &\\
\cline{2-3}
& Asymmetric tanh & 0.53520 &\\
\cline{1-3}
\multicolumn{2}{|c|}{Constant friction and temperature} & 0.3541 &\\
\hline
\end{tabular}
 \caption{For harmonically bound system with $k=1$.}
 \label{k1}
\end{table}

\section{Conclusion}
\label{conclusion}
In conclusion, we have revisited the celebrated BvL theorem from the perspective of fluctuations of in magnetic moment. We see that fluctuations are universal for both bounded and unbounded 
system when the friction coefficient is space dependent but the temperature is held fixed. However, for space dependent temperature, even though we recover BvL theorem in case of bounded system,
fluctuations in magnetic moment are non-universal. For a bound system, average entropy production ($\Delta S$) is zero due to the absence of probability currents in the system. For space
dependent friction, bounded system asymptotically reaches equilibrium and the probability distribution of entropy production is a delta function at $\Delta S=0$. However, this is not true for space dependent 
temperature. It exhibits finite fluctuation in probability distribution around $\Delta S=0$ \cite{acquino10,pal}.

\section{ACKNOWLEDGEMENTS}
A.M.J thanks Department of Science and Technology, India, for financial support (through J. C. Bose National Fellowship). A.M.J thanks N. Kumar and S. D. Mahanti for several useful discussions
 during the initial period of this work.

\end{document}